\RequirePackage{lineno}
\documentclass[aps,prd,twocolumn,tightenlines,superscriptaddress,showpacs,amsmath,amssymb]{revtex4}
\usepackage{graphicx} %
\usepackage{epstopdf} %
\usepackage{dcolumn} %
\usepackage{xcolor} %
\usepackage{amsmath,amssymb} %
\usepackage{hyperref} %
\usepackage[utf8]{inputenc} %
\usepackage[english]{babel} %
\usepackage{ifthen} %
\usepackage{orcidlink} %

\usepackage[noabbrev,capitalise]{cleveref} %

\usepackage{afterpage} %
\usepackage{rotating,booktabs,multirow} %

\graphicspath{{figures/}} %

\newboolean{articletitles}
\setboolean{articletitles}{true} %

\newcommand*\patchAmsMathEnvironmentForLineno[1]{%
\expandafter\let\csname old#1\expandafter\endcsname\csname #1\endcsname
\expandafter\let\csname oldend#1\expandafter\endcsname\csname
end#1\endcsname
 \renewenvironment{#1}%
   {\linenomath\csname old#1\endcsname}%
   {\csname oldend#1\endcsname\endlinenomath}%
}
\newcommand*\patchBothAmsMathEnvironmentsForLineno[1]{%
  \patchAmsMathEnvironmentForLineno{#1}%
  \patchAmsMathEnvironmentForLineno{#1*}%
}
\AtBeginDocument{%
\patchBothAmsMathEnvironmentsForLineno{equation}%
\patchBothAmsMathEnvironmentsForLineno{align}%
\patchBothAmsMathEnvironmentsForLineno{flalign}%
\patchBothAmsMathEnvironmentsForLineno{alignat}%
\patchBothAmsMathEnvironmentsForLineno{gather}%
\patchBothAmsMathEnvironmentsForLineno{multline}%
\patchBothAmsMathEnvironmentsForLineno{eqnarray}%
}

\RequirePackage{xspace}
\RequirePackage{relsize}

\def\babar{\mbox{\slshape B\kern-0.1em{\smaller A}\kern-0.1em
    B\kern-0.1em{\smaller A\kern-0.2em R}}\xspace}

\def\Ppi         {\ensuremath{\pi}\xspace}

\mathchardef\PDelta="7101
\mathchardef\PXi="7104
\mathchardef\PLambda="7103
\mathchardef\PSigma="7106
\mathchardef\POmega="710A
\mathchardef\PUpsilon="7107
                 
\def\PB      {\ensuremath{B}\xspace}                 
                 
\def\PD      {\ensuremath{D}\xspace}

\def\PK      {\ensuremath{K}\xspace}

\def\Pc      {\ensuremath{c}\xspace}                 
                 
\def\Pe      {\ensuremath{e}\xspace}

\def\Pi      {\ensuremath{i}\xspace}

\def\Pq      {\ensuremath{q}\xspace}                 
                 
\def\Ps      {\ensuremath{s}\xspace}

\def\epem       {\ensuremath{\Pe^+\Pe^-}\xspace}

\def\quark     {\ensuremath{\Pq}\xspace}
\def\quarkbar  {\ensuremath{\overline \quark}\xspace}
\def\qqbar     {\ensuremath{\quark\quarkbar}\xspace}

\def\squark    {\ensuremath{\Ps}\xspace}

\def\cquark    {\ensuremath{\Pc}\xspace}
\def\cquarkbar {\ensuremath{\overline \cquark}\xspace}
\def\ccbar     {\ensuremath{\cquark\cquarkbar}\xspace}

\def\pion  {\ensuremath{\Ppi}\xspace}

\def\pip   {\ensuremath{\pion^+}\xspace}
\def\pim   {\ensuremath{\pion^-}\xspace}

\def\kaon  {\ensuremath{\PK}\xspace}
\def\Kbar  {\kern 0.2em\overline{\kern -0.2em \PK}{}\xspace}%

\def\Kz    {\ensuremath{\kaon^0}\xspace}
\def\Kzb   {\ensuremath{\Kbar^0}\xspace}
\def\KzKzb {\ensuremath{\Kz \kern -0.16em \Kzb}\xspace}
\def\Kp    {\ensuremath{\kaon^+}\xspace}
\def\Km    {\ensuremath{\kaon^-}\xspace}

\def\KpKm  {\ensuremath{\Kp \kern -0.16em \Km}\xspace}
\def\KS    {\ensuremath{\kaon^0_{\rm\scriptscriptstyle S}}\xspace} 
\def\KL    {\ensuremath{\kaon^0_{\rm\scriptscriptstyle L}}\xspace}

\def\D       {\ensuremath{\PD}\xspace}
\def\Dbar    {\kern 0.2em\overline{\kern -0.2em \PD}{}\xspace}%

\def\Dz      {\ensuremath{\D^0}\xspace}
\def\Dzb     {\ensuremath{\Dbar^0}\xspace}
\def\DzDzb   {\ensuremath{\Dz {\kern -0.16em \Dzb}}\xspace}
\def\Dp      {\ensuremath{\D^+}\xspace}
\def\Dm      {\ensuremath{\D^-}\xspace}

\def\DpDm    {\ensuremath{\Dp {\kern -0.16em \Dm}}\xspace}

\def\Dstarp  {\ensuremath{\D^{*+}}\xspace}

\def\Dsp     {\ensuremath{\D^+_\squark}\xspace}

\def\Bbar    {\ensuremath{\kern 0.18em\overline{\kern -0.18em \PB}{}}\xspace}%

\def\Y#1S{\ensuremath{\PUpsilon{(#1S)}}\xspace}%

\def\Lbar {\ensuremath{\kern 0.1em\overline{\kern -0.1em\PLambda}}\xspace}

\def\to                 {\ensuremath{\rightarrow}\xspace}

\def\CP                {\ensuremath{C\!P}\xspace}

\newcommand{\tev}{\ensuremath{\mathrm{\,Te\kern -0.1em V}}\xspace}
\newcommand{\gev}{\ensuremath{\mathrm{\,Ge\kern -0.1em V}}\xspace}
\newcommand{\mev}{\ensuremath{\mathrm{\,Me\kern -0.1em V}}\xspace}
\newcommand{\kev}{\ensuremath{\mathrm{\,ke\kern -0.1em V}}\xspace}
\newcommand{\ev}{\ensuremath{\mathrm{\,e\kern -0.1em V}}\xspace}
\newcommand{\gevc}{\ensuremath{{\mathrm{\,Ge\kern -0.1em V\!/}c}}\xspace}
\newcommand{\mevc}{\ensuremath{{\mathrm{\,Me\kern -0.1em V\!/}c}}\xspace}
\newcommand{\gevcc}{\ensuremath{{\mathrm{\,Ge\kern -0.1em V\!/}c^2}}\xspace}
\newcommand{\gevgevcccc}{\ensuremath{{\mathrm{\,Ge\kern -0.1em V^2\!/}c^4}}\xspace}
\newcommand{\mevcc}{\ensuremath{{\mathrm{\,Me\kern -0.1em V\!/}c^2}}\xspace}

\def\cm   {\ensuremath{\rm \,cm}\xspace}

\def\invfb   {\ensuremath{\mbox{\,fb}^{-1}}\xspace}

\def\gsim{{~\raise.15em\hbox{$>$}\kern-.85em
          \lower.35em\hbox{$\sim$}~}\xspace}
\def\lsim{{~\raise.15em\hbox{$<$}\kern-.85em
          \lower.35em\hbox{$\sim$}~}\xspace}

\newcommand{\ie}{\mbox{\itshape i.e.}}

\newcommand{\Acp}{\ensuremath{A_{\CP}}\xspace}

\newcommand{\mD}{\ensuremath{m(\KS\KS)}\xspace}

\newcommand{\Smin}{\ensuremath{S_\text{min}(\KS)}\xspace}

\newcommand{\DzToKSKS}{\ensuremath{\Dz\to\KS\KS}\xspace}

\newcommand{\resBelle}{2.5}
\newcommand{\statBelle}{2.7}
\newcommand{\systBelle}{0.4}

\newcommand{\resBelleII}{-0.1}
\newcommand{\statBelleII}{3.0}
\newcommand{\systBelleII}{0.3}

\newcommand{\resValue}{1.3}
\newcommand{\resStat}{2.0}
\newcommand{\resSyst}{0.2}

\newcommand{\resComb}{-0.6}
\newcommand{\statComb}{1.1}
\newcommand{\systComb}{0.1}
\makeatletter%
\begin{document}
% inside_import 
% before 
% ignored 
% args [width=3cm]
% full_filename belle2-logo.pdf
% after \vspace*{-1.9cm}
\includegraphics[width=3cm]{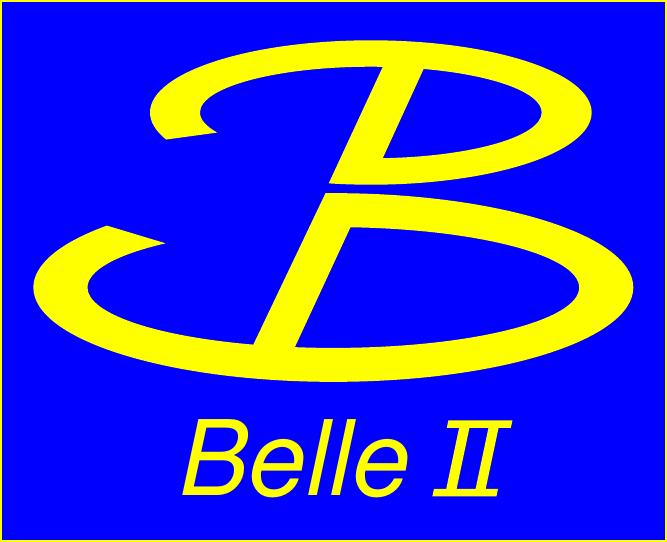}\vspace*{-1.9cm}

\begin{flushright}
Belle II Preprint 2025-006\\
KEK Preprint 2025-4
\end{flushright}\vspace{1.5cm}

\title{%
% start input ./title.tex
%
{\LARGE\bfseries\boldmath Measurement of the time-integrated \CP asymmetry in $\Dz\to\KS\KS$ decays using opposite-side flavor tagging at Belle and Belle II}
 % end input ./title.tex
 %
}
%
% start input ./authors/pub079-orcid.tex
%
%
%
%
%
  \author{I.~Adachi\,\orcidlink{0000-0003-2287-0173}} %
  \author{Y.~Ahn\,\orcidlink{0000-0001-6820-0576}} %
  \author{N.~Akopov\,\orcidlink{0000-0002-4425-2096}} %
  \author{S.~Alghamdi\,\orcidlink{0000-0001-7609-112X}} %
  \author{M.~Alhakami\,\orcidlink{0000-0002-2234-8628}} %
  \author{A.~Aloisio\,\orcidlink{0000-0002-3883-6693}} %
  \author{N.~Althubiti\,\orcidlink{0000-0003-1513-0409}} %
  \author{K.~Amos\,\orcidlink{0000-0003-1757-5620}} %
  \author{M.~Angelsmark\,\orcidlink{0000-0003-4745-1020}} %
  \author{N.~Anh~Ky\,\orcidlink{0000-0003-0471-197X}} %
% ...
%
\collaboration{The Belle and Belle II Collaborations}
 % end input ./authors/pub079-orcid.tex
 
\begin{abstract}
%
% start input ./abstract.tex
%
\noindent We measure the time-integrated \CP asymmetry in \DzToKSKS decays reconstructed in $e^+e^-\to\ccbar$ events collected by the Belle and Belle~II experiments. The corresponding data samples have integrated luminosities of 980 and 428\invfb, respectively. To infer the flavor of the \Dz meson, we exploit the correlation between the flavor of the reconstructed decay and the electric charges of particles reconstructed in the rest of the $\epem\to\ccbar$ event. This results in a sample which is independent from any other previously used at Belle or Belle~II. The result, $\Acp(\DzToKSKS) = (\resValue\pm\resStat\pm\resSyst)\%$, where the first uncertainty is statistical and the second systematic, is consistent with previous determinations and with \CP symmetry. % end input ./abstract.tex
 \end{abstract}

\maketitle
%
% start input ./body.tex
%
\section{Introduction}
Decays of charm hadrons offer a unique avenue for exploring flavor and charge-parity (\CP) violation in the sector of up-type quarks, which is complementary to the searches performed with strange and beauty hadrons. The dynamics of charm decays is complicated by the presence of non-perturbative QCD effects that are difficult to calculate, making it a test case for both the electroweak and strong interactions. After being first observed in 2019~\cite{Aaij:2019kcg}, \CP violation in charm decays has gained renewed attention. The nature of the observed \CP violation has yet to be fully understood, and could be due to enhanced non-perturbative QCD effects or to physics beyond the standard model
~\cite{Chala:2019fdb,Dery:2019ysp,Calibbi:2019bay,Grossman:2019xcj,Cheng:2019ggx,Buras:2021rdg,Schacht:2021jaz,Schacht:2022kuj,Bediaga:2022sxw,Bause:2022jes,Pich:2023kim,Gavrilova:2023fzy}. Flavor and isospin symmetries can be used to relate measurements from different decay modes, helping to constrain non-perturbative QCD effects and identify possible new physics contributions~\cite{Hiller:2012xm,Grossman:2019xcj,Buccella:2019kpn}. Hence, searches in additional channels and improved measurements of \CP asymmetries in already explored decay modes are important.

In this paper, we report a measurement of the time-integrated \CP asymmetry in \DzToKSKS decays using a combination of Belle and Belle II data, which have integrated luminosities of 980 and 428\invfb, respectively~\cite{Brodzicka:2012jm,lumi}. The time-integrated \CP asymmetry is defined as
\begin{multline}
\Acp(\DzToKSKS) = \\
\frac{\Gamma(\Dz\to\KS\KS)-\Gamma(\Dzb\to\KS\KS)}{\Gamma(\Dz\to\KS\KS)+\Gamma(\Dzb\to\KS\KS)}\,,
\end{multline}
where $\Gamma$ indicates the decay rate integrated over decay time, which includes effects due to \Dz-\Dzb mixing. Experimental measurements of $\Acp(\DzToKSKS)$~\cite{CLEO:2000opx,LHCb:2015ope,LHCb:2021rdn,CMS:2024hsv,Belle:2024vho} are consistent with \CP symmetry within about 1\%, putting them at the upper limit of that predicted by the standard model and in the range where new-physics contributions might be seen~\cite{Buccella:2019kpn,Brod:2011re,Nierste:2015zra,PhysRevD.86.036012,PhysRevD.100.093002}.

An important part in the measurement of $\Acp(\DzToKSKS)$ is the ability to determine the production flavor of the neutral \D meson, which is referred to as ``tagging.'' All measurements performed so far use neutral \D decays originating from the strong-interaction decay $\Dstarp\to\Dz\pip$, where the charge of the accompanying pion can be used for tagging. (Charge-conjugate modes are implied throughout the paper, unless stated otherwise.) In this measurement, instead, we use the charm flavor tagger (CFT) described in Ref.~\cite{Belle-II:2023vra}. The CFT identifies the flavor of a reconstructed neutral \D meson by exploiting correlations with the electric charges of particles reconstructed in the rest of the $\epem\to\ccbar$ event. These include those originating from the decay of the other charm hadron produced in the event, as well as those possibly produced in association with the reconstructed \D meson, such as in $\Dstarp\to\Dz\pip$ decays. To avoid correlations with the recently published \Dstarp-tagged measurement of Ref.~\cite{Belle:2024vho}, we use an independent dataset where all the candidates previously used are removed. As a consequence, the CFT in this measurement acts as an ``opposite-side'' tagger using information from the other charm hadron in the event. The CFT is calibrated on large data samples of $\Dz\to\Km\pip$ decays, following the procedure described in Ref.~\cite{Belle-II:2023vra}. The time-integrated \CP asymmetry is then determined from an unbinned maximum-likelihood fit to the two-dimensional distribution of the $\KS\KS$ mass and the CFT output. To avoid potential bias, an arbitrary and undisclosed offset was added to the measured value of $\Acp(\DzToKSKS)$ when fitting to the data. The offset remained undisclosed until the entire analysis procedure was completed and the determination of all uncertainties was finalized.

The paper is organized as follows. \Cref{sec:detector} provides an overview of the Belle and Belle II detectors. \cref{sec:simulation} details the simulation samples used in the measurement. The reconstruction and selection of the signal \DzToKSKS decays is presented in \cref{sec:selection}. Determination of the asymmetry is covered in \cref{sec:fit}, followed by a discussion of the systematic uncertainties affecting the measurement in \cref{sec:systematics}. The final results, and a combination with the \Dstarp-tagged measurement of Ref.~\cite{Belle:2024vho}, are presented in \cref{sec:results}.

\section{Belle and Belle II detectors\label{sec:detector}}
The Belle experiment~\cite{belle_detector,Brodzicka:2012jm} operated at the KEKB asymmetric-energy $\epem$ collider~\cite{kekb,Abe:2013kxa} between 1999 and 2010. The detector consisted of a large-solid-angle spectrometer, which included a double-sided silicon-strip vertex detector, a 50-layer central drift chamber, an array of aerogel threshold Cherenkov counters, a barrel-like arrangement of time-of-flight scintillation counters, and an electromagnetic calorimeter comprised of CsI(Tl) crystals. All subdetectors were located inside a superconducting solenoid coil that provided a 1.5 T magnetic field. An iron flux-return yoke, placed outside the coil, was instrumented with resistive-plate chambers to detect \KL mesons and identify muons.  Two inner detector configurations were used: a 2.0\cm radius beam pipe and a three-layer silicon vertex detector; and, from October 2003, a 1.5\cm radius beam pipe, a four-layer silicon vertex detector, and a small-inner-cell drift chamber~\cite{Natkaniec:2006rv}.

The Belle II detector~\cite{b2tech,Kou:2018nap} is an upgrade with several new subdetectors designed to handle the significantly larger beam-related backgrounds of the new SuperKEKB $\epem$ collider~\cite{Akai:2018mbz}. It consists of a silicon vertex detector wrapped around a 1\cm radius beam pipe and comprising two inner layers of pixel detectors and four outer layers of double-sided strip detectors, a 56-layer central drift chamber, a time-of-propagation detector, an aerogel ring-imaging Cherenkov detector, and an electromagnetic calorimeter, all located inside the same solenoid as used for Belle. A flux return outside the solenoid is instrumented with resistive-plate chambers, plastic scintillator modules, and an upgraded read-out system to detect muons and \KL mesons. For the data used in this paper, collected between 2019 and 2022, only part of the second layer of the pixel detector, covering 15\% of the azimuthal angle, was installed.

For both experiments, the $z$ axis of the laboratory frame is defined as the central axis of the solenoid, with its positive direction determined by the direction of the electron beam.

\section{Simulation\label{sec:simulation}}
We use simulated event samples to identify sources of background, optimize selection criteria, and validate the analysis procedure. We generate $\epem\to\Upsilon(nS)$ ($n=4,5$) events and simulate particle decays with \textsc{EvtGen}~\cite{Lange:2001uf}; we generate continuum $\epem\to\qqbar$ (where $q$ is a $u$, $d$, $c$, or $s$ quark) with \textsc{Pythia6}~\cite{Sjostrand:2006za} for Belle, and with \textsc{KKMC}~\cite{Jadach:1999vf} and \textsc{Pythia8}~\cite{Sjostrand:2014zea} for Belle~II; we simulate final-state radiation with \textsc{Photos}~\cite{Barberio:1990ms,Barberio:1993qi}; we simulate detector response using \textsc{Geant3}~\cite{Brun:1073159} for Belle and \textsc{Geant4}~\cite{Agostinelli:2002hh} for Belle II. Beam backgrounds are taken into account by overlaying random trigger data.

\section{Reconstruction and event selection\label{sec:selection}}
We use the Belle II analysis software framework (basf2) to reconstruct both Belle and Belle~II data~\cite{Kuhr:2018lps,basf2-zenodo}. The Belle data are converted to the Belle II format for basf2 compatibility using the B2BII framework~\cite{Gelb:2018agf}. 

Events are selected by a trigger based on either the total energy deposited in the electromagnetic calorimeter or the number of charged-particle tracks reconstructed in the central drift chamber. The efficiency of the trigger is found to be close to 100\% for signal decays.

Candidate $\KS\to\pip\pim$ decays are reconstructed from combinations of oppositely charged particles that are constrained to originate from a common vertex. These particles are assumed to be pions and the resulting dipion mass is required to be in the $[0.45,0.55]\gevcc$ range. Pairs of \KS candidates are combined to form candidate \DzToKSKS decays. We perform a kinematic fit~\cite{Krohn:2019dlq} to the \Dz candidates by constraining their momentum directions to point back to the measured position of the beam interaction point, and constraining the masses of the two \KS candidates to the nominal \KS mass~\cite{pdg}. Only candidates whose kinematic fits converge are retained for further analysis. The mass of the \Dz candidate, \mD, is required to be in the range $[1.84,2.00]\gevcc$ to exclude partially reconstructed $\Dsp\to\KS\KS\pip$ decays, which peak at lower mass values. The \Dz momentum in the $\epem$ center-of-mass system is required to be greater than 2.2\gevc to suppress candidates in which the \Dz meson arises from the decay of a $B$ meson, as the CFT is trained and calibrated for $\epem\to\ccbar$ events. Candidates that are also reconstructed in the \Dstarp-tagged analysis of Ref.~\cite{Belle:2024vho} are removed.

To suppress $\Dz\to\KS\pip\pim$ decays, we use the large distance ($L$) between the \KS and \Dz decay vertices resulting from the long \KS lifetime. We introduce the variable $\Smin=\log\!\left[\min\!\left(L_1/\sigma_{L_1}, L_2/\sigma_{L_2} \right) \right]$, where $L_{1(2)}$ and $\sigma_{L_{1(2)}}$ are the distance and its uncertainty for the first (second) \KS candidate, respectively~\cite{Belle:2024vho}. Candidates satisfying  $\min\!\left(L_1/\sigma_{L_1}, L_2/\sigma_{L_2} \right)\leq 0$ are removed. We require \Smin to be larger than 1.75 and 2.05 for the Belle and Belle II samples, respectively.

\begin{figure}[ht]
\centering
% inside_import 
% before 
% ignored 
% args [width=.4\textwidth]
% full_filename figures/BDTplots_Belle.pdf
% after \\
\includegraphics[width=.4\textwidth]{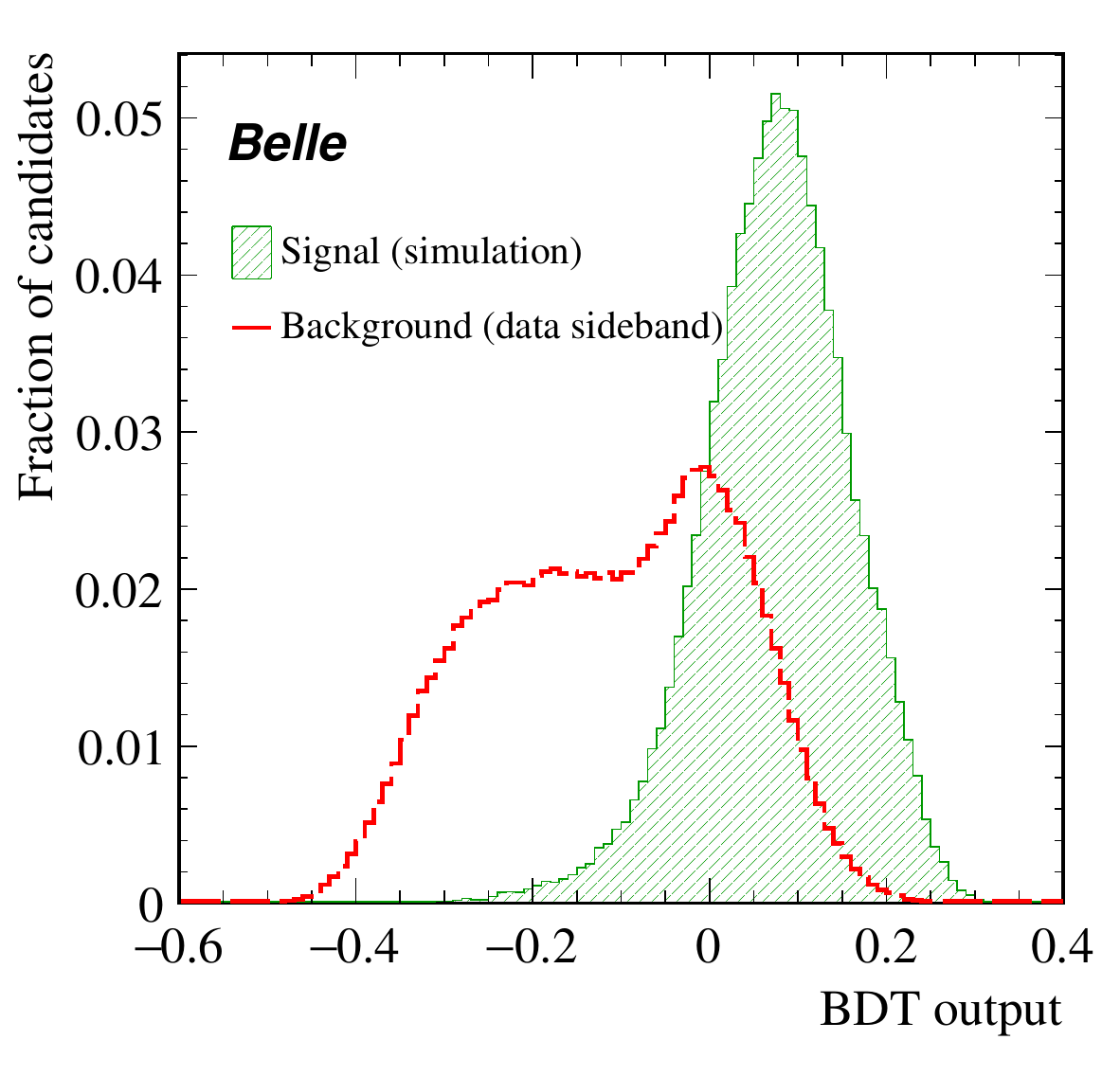}\\
% inside_import 
% before 
% ignored 
% args [width=.4\textwidth]
% full_filename figures/BDTplots_B2.pdf
% after \\
\includegraphics[width=.4\textwidth]{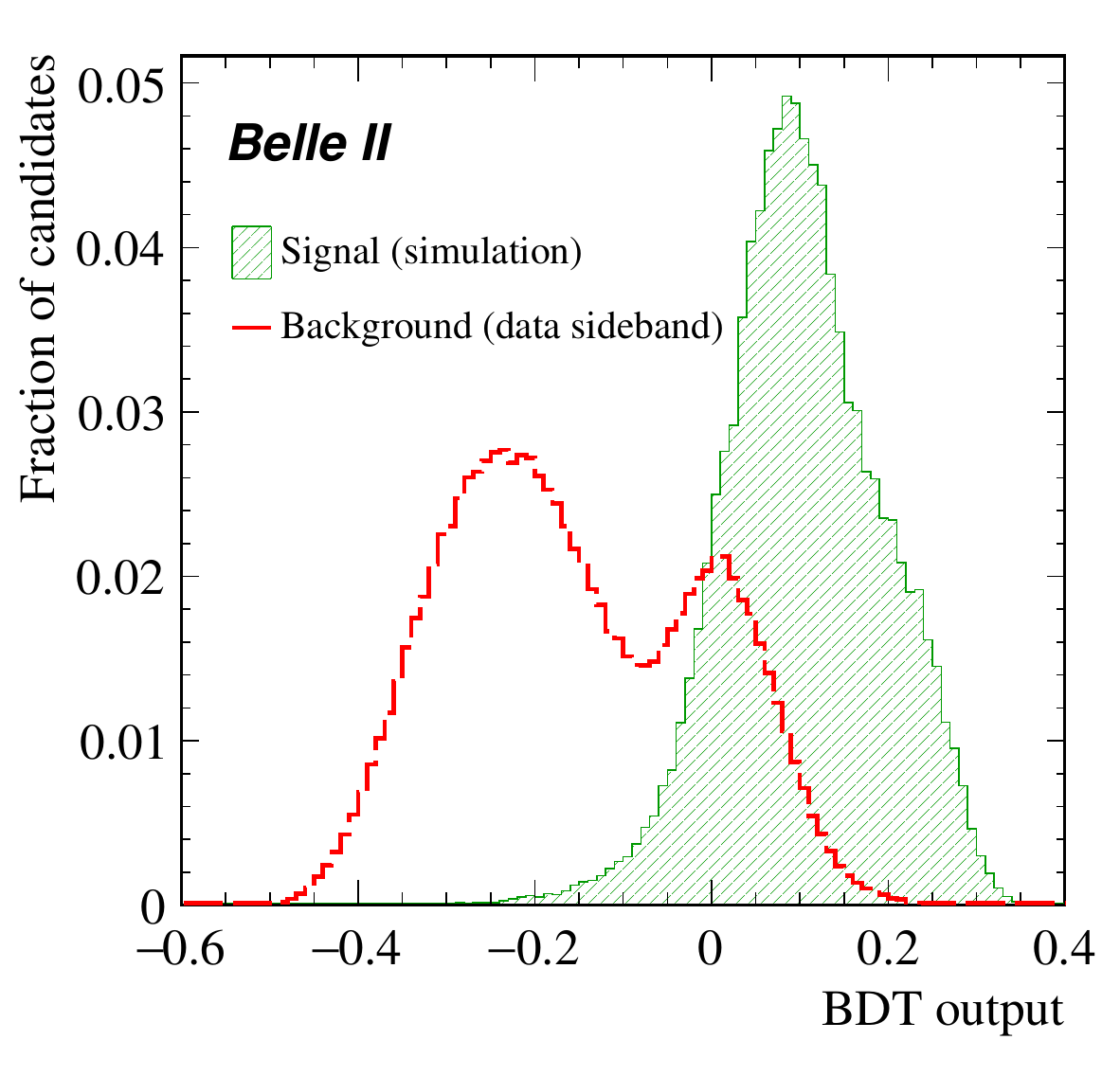}\\
\caption{Distributions of the BDT output for the training samples of (green) signal and (red) background candidates in (top) Belle and (bottom) Belle II.\label{fig:BDT_plots}}
\end{figure}

Combinatorial background from two unrelated \KS candidates is suppressed using the output of a boosted decision tree (BDT) trained to discriminate such background from signal decays~\cite{Hocker:2007ht,TMVA2007}. The BDT is trained using simulated \DzToKSKS decays as signal, and data candidates populating the \mD sideband $[1.90,2.00]\gevcc$ as background. The input variables to the BDT are selected to effectively separate the signal from the background, while minimizing any correlations with $m(\KS\KS)$ and \Smin. The input variables are the logarithm of the minimum of the transverse impact parameters of the four final-state pions, the logarithm of the minimum of the longitudinal impact parameters of the four final-state pions, the maximum of the momenta of the final-state pions of the \KS candidate with lower momentum, the minimum of the momenta of the final-state pions of the \KS candidate with lower momentum, the absolute value of the polar angle difference between the final-state pions of the \KS candidate with lower momentum, the invariant masses of the two \KS candidates, the scalar sum of the momenta of the two \KS candidates, the absolute value of the asymmetry between the momenta of the two \KS candidates, the flight-distance of the \Dz candidate divided by its uncertainty, and the logarithm of the $\chi^2$ probability of the vertex fit of the \Dz candidate. The BDT output distributions for the training samples of signal and background candidates are shown in \cref{fig:BDT_plots}. The background distribution shows a two-peak structure, arising from candidates formed with either correctly reconstructed \KS candidates or random pion combinations. Due to the additional pixel detector layers and the improved reconstruction performance, the Belle II sample has a better signal-to-background separation as compared to Belle. We require the BDT output to exceed 0.037 and 0.075 for Belle and Belle II, respectively.

The requirements on \Smin and on the BDT output are optimized simultaneously by maximizing the quantity $S/\sqrt{S+B}$, where $S$ and $B$ are simulated signal and background yields in a $3\sigma$ \mD range around the signal peak (\ie, the signal region shown in \cref{fig:data_fit}), while keeping the residual $\Dz\to\KS\pip\pim$ background below 3\% of the \DzToKSKS yield. The 3\% threshold ensures that the measured asymmetry cannot be biased by more than about one tenth of the expected statistical uncertainty, as discussed in \cref{sec:systematics}. The optimization is done separately for the Belle and Belle~II samples. The optimized requirements have signal efficiencies of 55\% in Belle and 58\% in Belle~II, and background rejections of 87\% for Belle and 94\% for Belle~II.

We use the CFT to predict the flavor $q$ of the selected candidates ($q=+1$ for \Dz, $q=-1$ for \Dzb). The tagger also outputs an associated dilution factor $r$, which is related to the per-candidate mistag probability $\omega$ by $r = 1-2\omega$. We remove untagged candidates, \ie, candidates for which the CFT does not produce any flavor prediction ($qr=0$), which amount to 0.2\% and 0.6\% of the total in Belle and Belle II, respectively. We accept all candidates in events where multiple \Dz candidates are reconstructed, which amount to 1\% and 0.3\% of the total in Belle and Belle II, respectively.

\section{Determination of the \CP asymmetry\label{sec:fit}}
The \CP asymmetry is determined using an unbinned maximum-likelihood fit to the \mD and $r$ distributions of the two flavors $q=\pm1$. The \mD distribution discriminates the \DzToKSKS signal component from a smooth combinatorial background. The $qr$ distribution determines the asymmetry between \Dz and \Dzb candidates. The two-dimensional probability density functions (PDFs) of each component can be factorized into the product of one-dimensional PDFs. The signal \mD PDF, $P_{\rm s}(m)$, is modeled using the sum of two Gaussian distributions. The background PDF, $P_{\rm b}(m)$, is modeled with an exponential distribution. We model the distributions of $r$, $P_{\rm s,b}(r)$, using histogram templates extracted directly from the data. For background, we use candidates populating the \mD sideband $[1.91,2.00]\gevcc$; for signal, we subtract the background distribution from that of the candidates in the \mD signal region. Simulation shows that the sideband data describe well the $r$ distribution of the background candidates in the signal region.

The symbolic expression of the PDF of a single candidate is
\begin{multline}\label{eq:full_pdf}
P(m, q, r|\Acp, A_{\rm b}, ...) = f_{\rm b}(1+qrA_{\rm b})P_{\rm b}(m|...)P_{\rm b}(r)\\
+(1-f_{\rm b})\left[1+q d(r|...)\Acp+q\Delta_d(r|...)\right]P_{\rm s}(m|...)P_{\rm s}(r)\,,
\end{multline}
where $f_{\rm b}$ is the fraction of background candidates, $A_{\rm b}$ is the observed background asymmetry, $d$ and $\Delta_d$ are third-order polynomials used to calibrate the per-candidate dilution, and the ellipses ($...$) indicates other fit parameters omitted here for brevity. All parameters are floated in the fit together with the signal asymmetry \Acp.

\begin{figure}[ht]
\centering
% inside_import 
% before 
% ignored 
% args [width=.4\textwidth]
% full_filename figures/calib_b1_data_nodst.pdf
% after \\
\includegraphics[width=.4\textwidth]{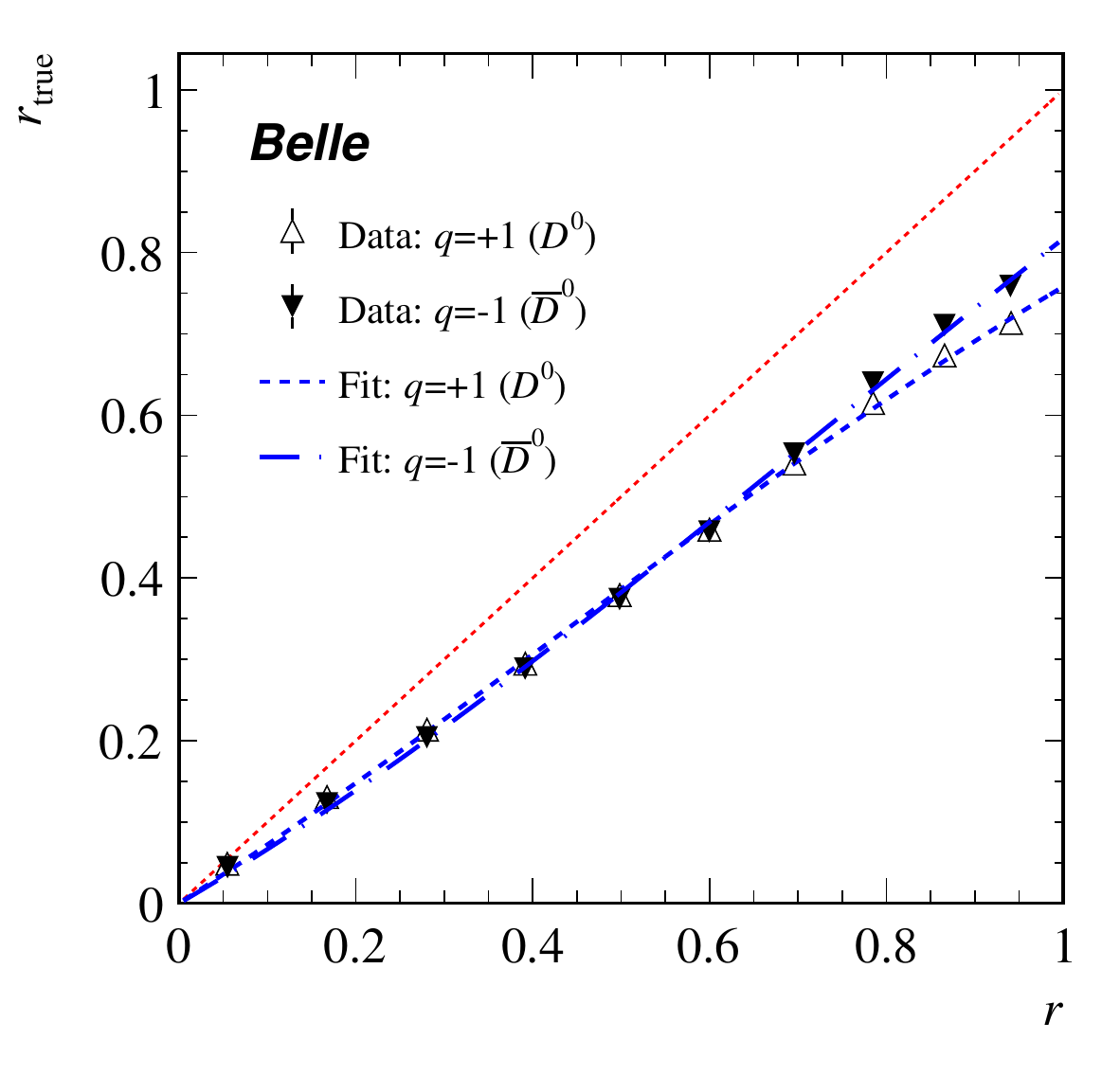}\\
% inside_import 
% before 
% ignored 
% args [width=.4\textwidth]
% full_filename figures/calib_b2_data_nodst.pdf
% after \\
\includegraphics[width=.4\textwidth]{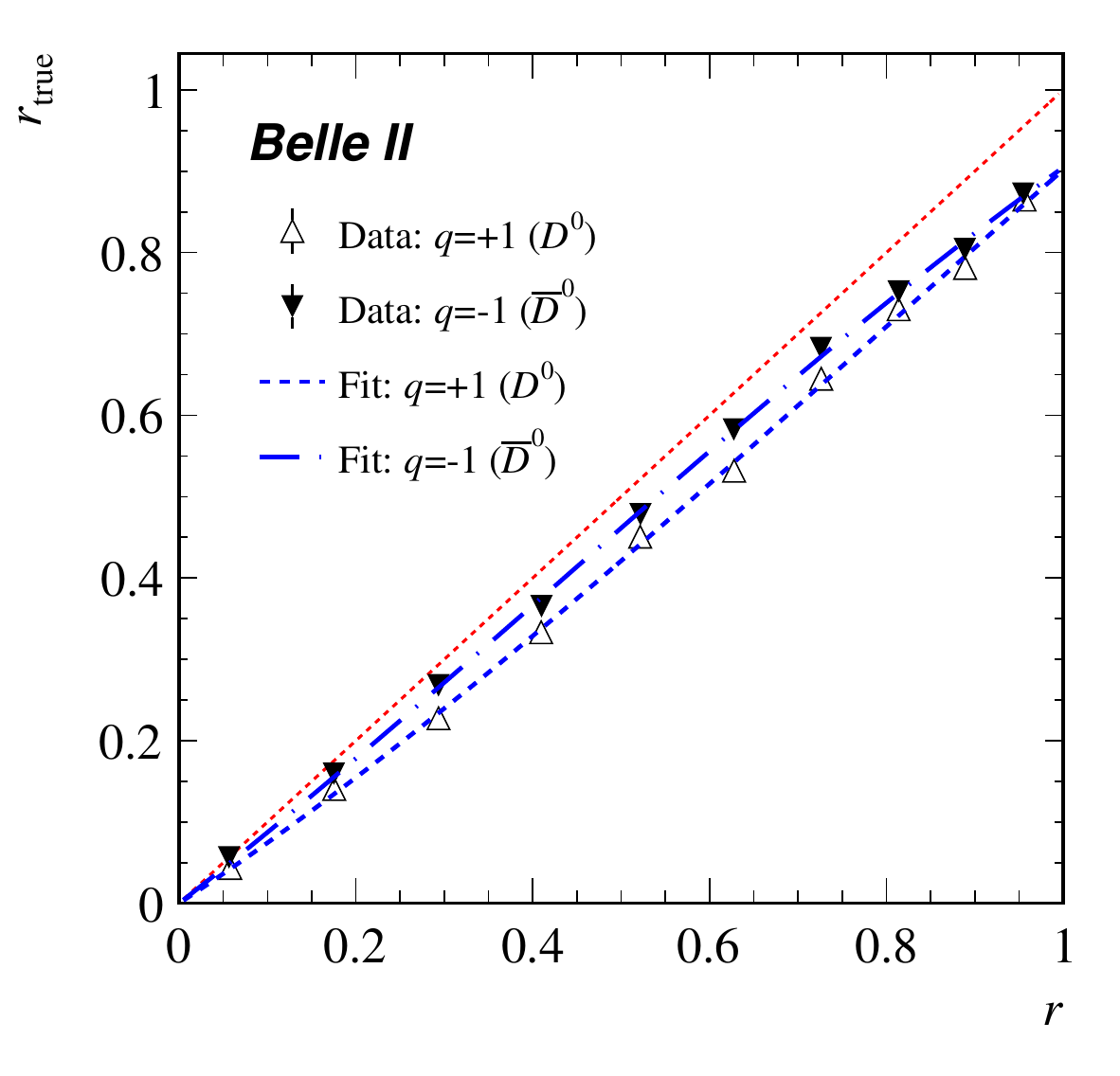}\\
\caption{True dilution as a function of the predicted dilution for $\Dz\to\Km\pip$ and $\Dzb\to\Kp\pim$ decays in (top) Belle and (bottom) Belle II data with projections of the calibration fit overlaid. The bisector of the plane (red dotted line) represents the expected relation for perfectly calibrated predicted dilution.\label{fig:calibration_plots}}
\end{figure}

The true dilution, $r^{\rm true}$, is expressed as a function of the predicted flavor $q$ and per-candidate dilution $r$ using
\begin{multline}\label{eq:calibration}
r^{\rm true}(q,r|p_1,p_2,p_3,\Delta_{p_1},\Delta_{p_2},\Delta_{p_3}) = d(r|p_1,p_2,p_3) \\
+ q\Delta_d(r|\Delta_{p_1},\Delta_{p_2},\Delta_{p_3})\,,
\end{multline}
where the coefficients of the polynomials ($p_1,...,\Delta_{p_1},...$) are determined by comparing predicted and true dilutions in high-yield data samples of $\Dz\to\Km\pip$ decays, where the neutral \D flavor is inferred from the charge of the final-state kaon~\cite{Belle-II:2023vra} (\cref{fig:calibration_plots}). The tagging power (or effective tagging efficiency), computed from the calibrated per-candidate dilution, is $(23.52\pm0.01)\%$ in Belle and $(32.71\pm0.05)\%$ in Belle II. The tagging power is lower than reported in Ref.~\cite{Belle-II:2023vra} due to the removal of the same-side \Dstarp-tagged candidates. To account for the uncertainties in the calibration parameters, a term constructed from their covariance matrix is included in the likelihood, such that the parameters are Gaussian constrained to their measured values. Thus, the systematic uncertainty associated with the knowledge of the calibration parameters is already taken into account in the statistical uncertainty returned by the fit.

Fits to the simulation and to pseudoexperiments generated by sampling from the PDF show no evidence of a bias in the determinations of the signal yield and asymmetry, nor in their uncertainties.

\begin{figure*}[ht]
\centering
% inside_import 
% before 
% ignored 
% args [width=.4\textwidth]
% full_filename figures/fitDatabelle_m.pdf
% after \hfil
\includegraphics[width=.4\textwidth]{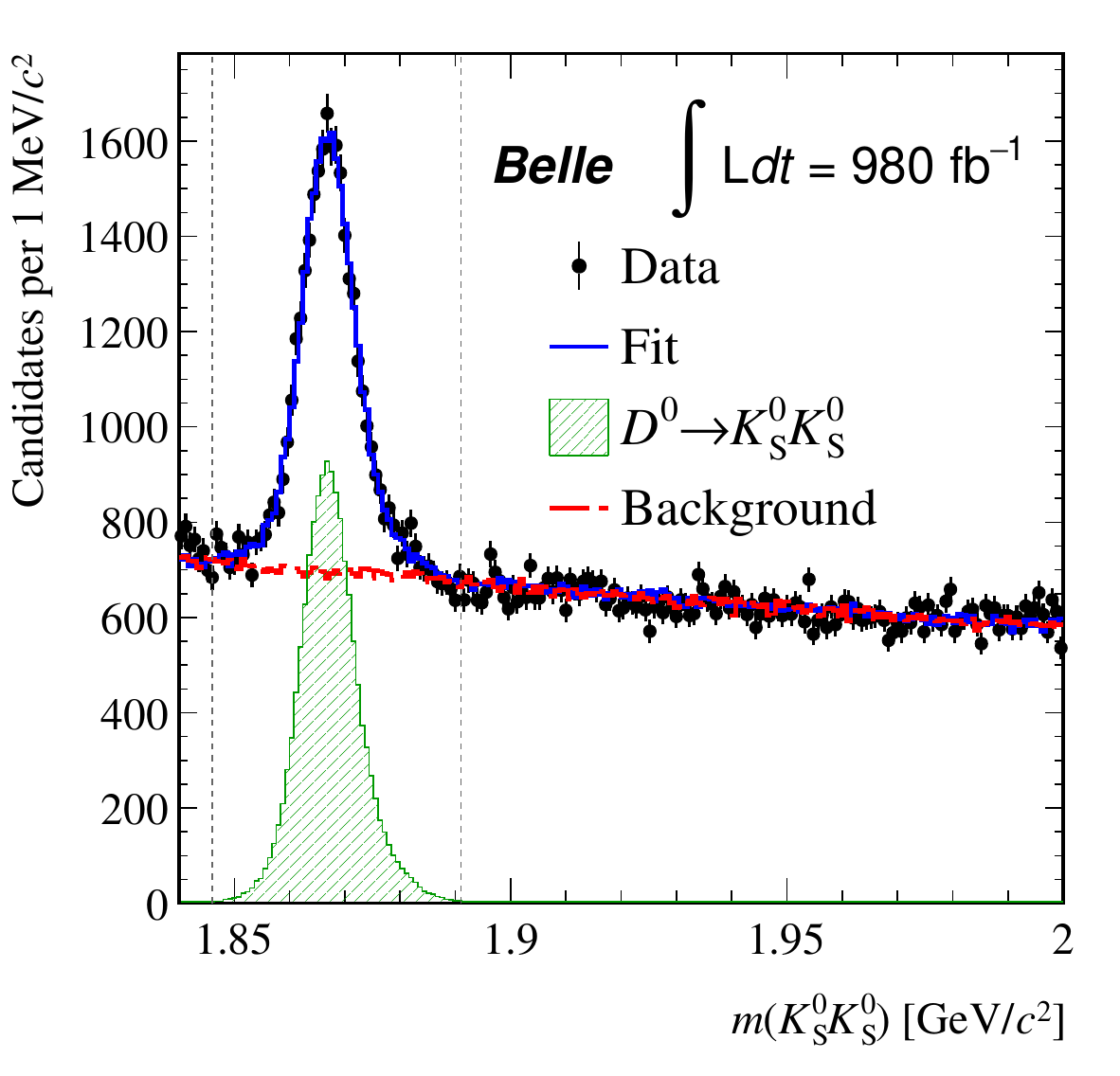}\hfil
% inside_import 
% before 
% ignored 
% args [width=.4\textwidth]
% full_filename figures/fitDatabelle_qr_sigEnhanced.pdf
% after \\
\includegraphics[width=.4\textwidth]{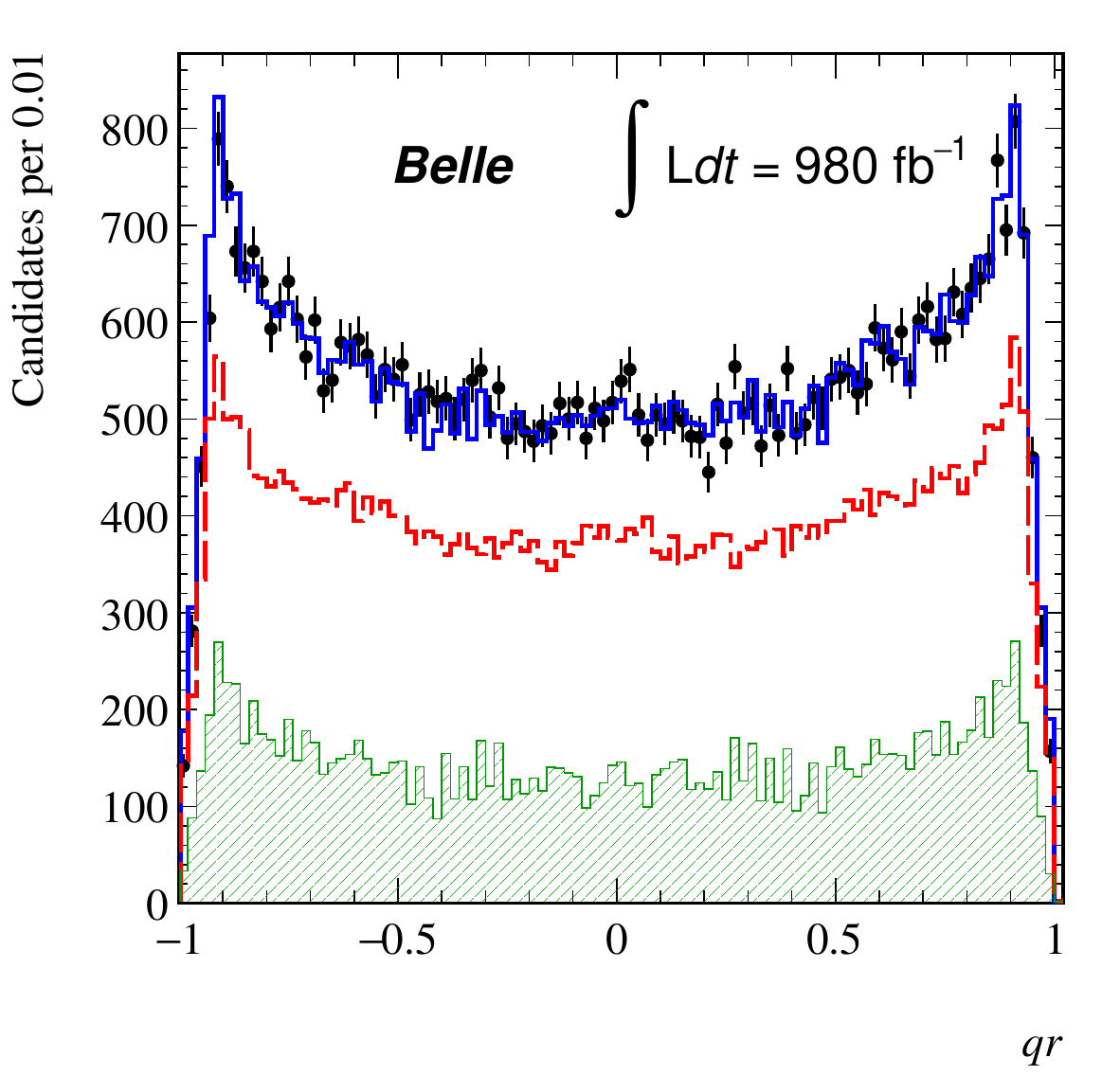}\\
% inside_import 
% before 
% ignored 
% args [width=.4\textwidth]
% full_filename figures/fitDatab2_m.pdf
% after \hfil
\includegraphics[width=.4\textwidth]{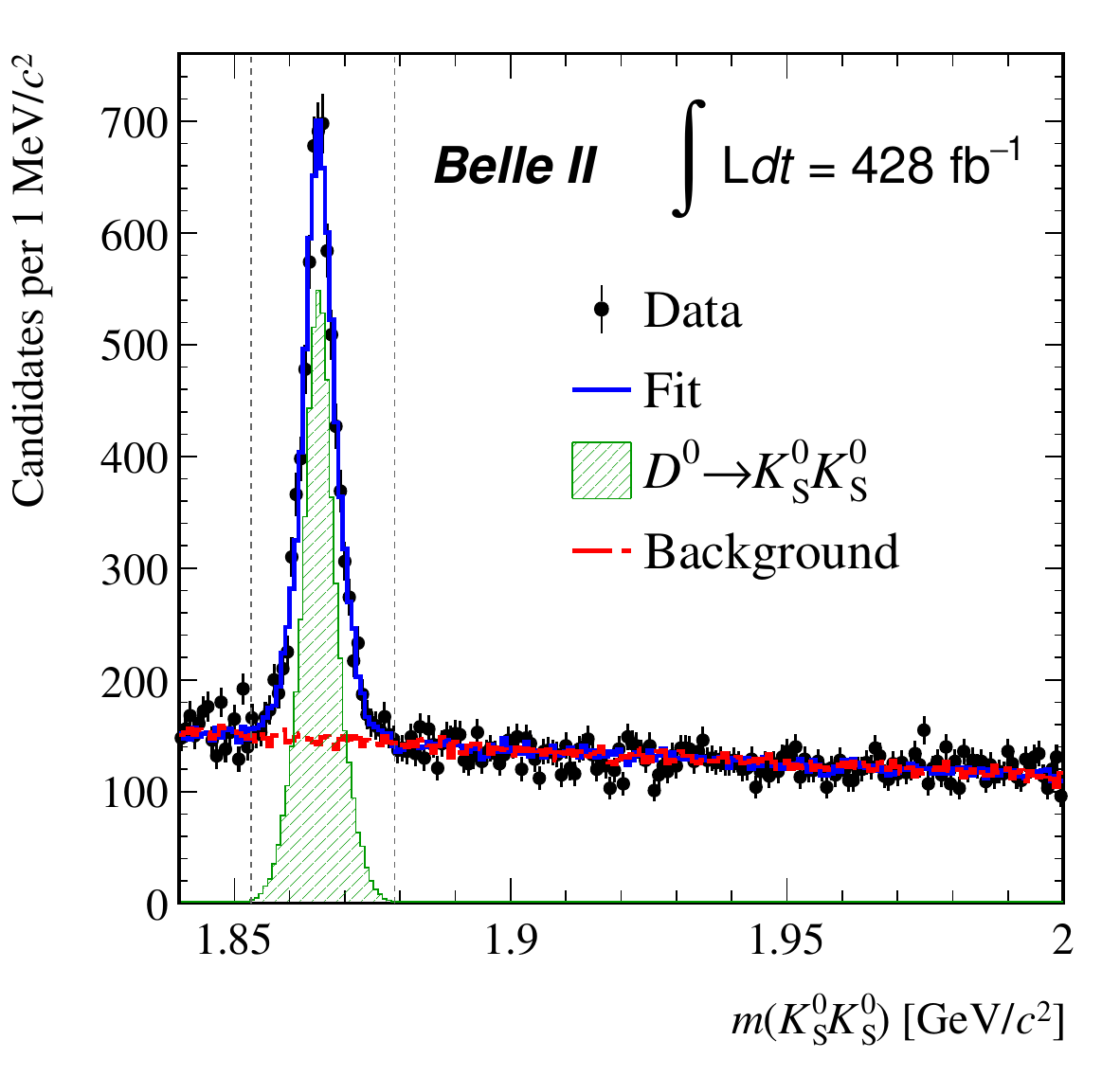}\hfil
% inside_import 
% before 
% ignored 
% args [width=.4\textwidth]
% full_filename figures/fitDatab2_qr_sigEnhanced.pdf
% after \\
\includegraphics[width=.4\textwidth]{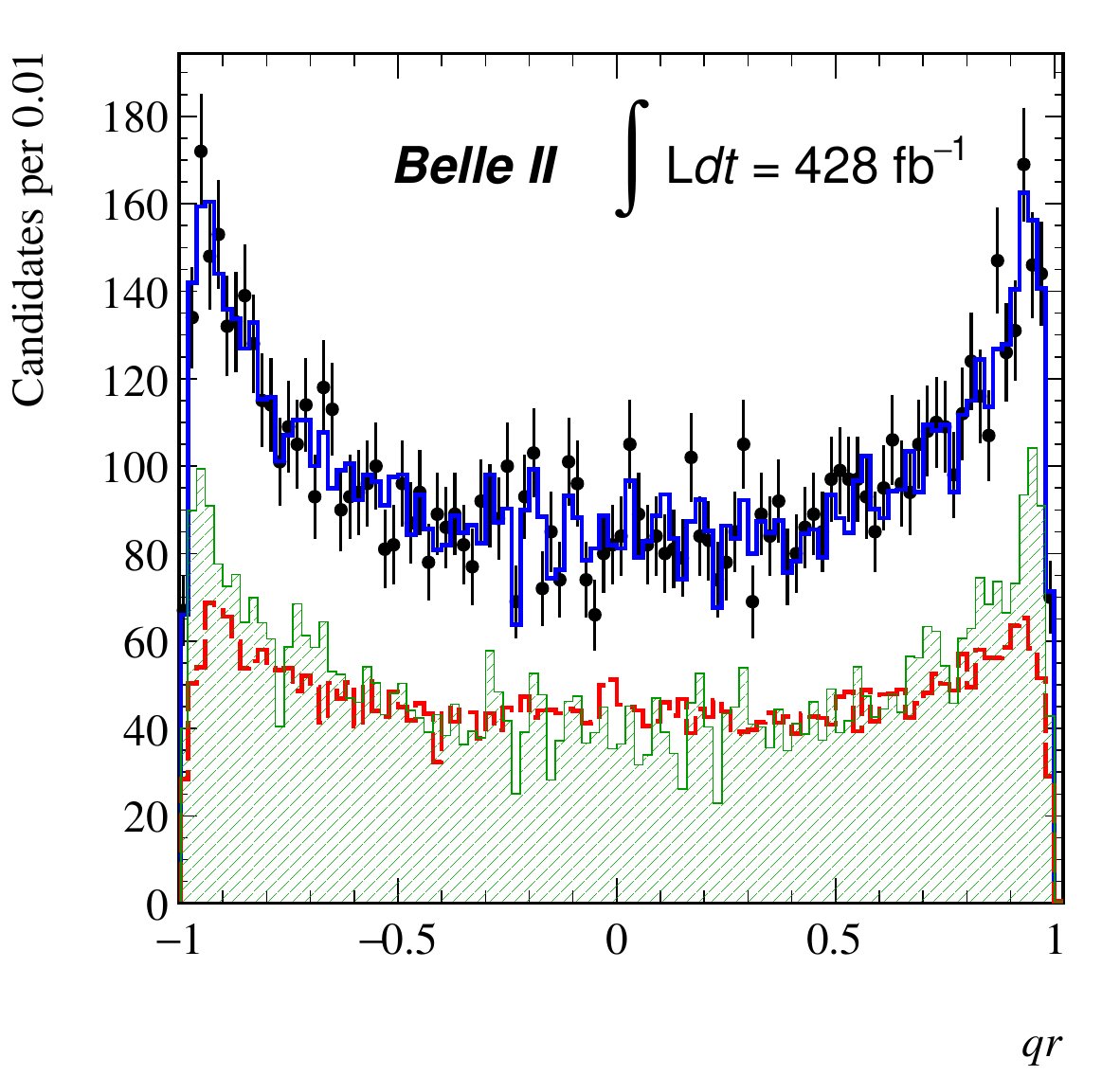}\\
\caption{Distributions of (left) \mD and (right) $qr$ for \DzToKSKS candidates in (top) Belle and (bottom) Belle II data, with fit projections overlaid. The $qr$ distributions are only for candidates in the \mD signal regions indicated by the vertical lines. %
\label{fig:data_fit}}
\end{figure*}

We perform independent fits to Belle and Belle II data. The fit model describes the data well, as shown in \cref{fig:data_fit}. The measured signal yields are $14490\pm340$ in Belle, and $5180\pm120$ in Belle II. The asymmetry is measured to be $(\resBelle\pm\statBelle)\%$ in Belle, and $(\resBelleII\pm\statBelleII)\%$ in Belle II. The uncertainties are statistical only. The two results are in agreement.

\section{Systematic uncertainties\label{sec:systematics}}
We consider the following sources of systematic uncertainties: fit modeling, residual contamination from $\Dz\to\KS\pip\pim$ decays, and effects due to the forward-backward asymmetry in $\epem\to\ccbar$ production.

We estimate the first using pseudoexperiments generated with the default fit model, and fitted with alternative models derived from data where one of the fit shapes is changed. As alternative models for the mass shapes we use a Johnson $S_U$ distribution~\cite{johnson} for signal and a second-order polynomial for background. For the $r$ distributions, we fill alternative histogram templates by varying the definition of the mass sideband. The alternative models give an equally good description of the data as the default models. The observed average differences between measured and generated asymmetries, 0.35\% for Belle and 0.10\% for Belle II, are assigned as systematic uncertainties.

The residual $\Dz\to\KS\pip\pim$ background, which is indistinguishable from the signal in \mD, is neglected in the fit and counted as part of the signal component. This introduces a bias on the measured asymmetry, which can be estimated as the product of the contamination fraction and the difference between the \CP asymmetries in $\Dz\to\KS\pip\pim$ and $\Dz\to\KS\KS$ decays. The contamination fraction is estimated in simulation to be 2.5\% for Belle, and 2.3\% for Belle II. Given that $\Acp(\Dz\to\KS\pip\pim)-\Acp(\Dz\to\KS\KS) < 10\%$~\cite{pdg}, the bias can be conservatively bounded to be smaller than 0.25\% for Belle, and 0.23\% for Belle II. These values are assigned as systematic uncertainties due to neglecting $\Dz\to\KS\pip\pim$ contamination, and are conservative enough to also cover possible detection asymmetries in the $\Dz\to\KS\pip\pim$ decay. 

In $\epem\to\ccbar$ events, charmed hadrons are produced with a forward-backward asymmetry due to $\gamma$-$Z^{0}$ interference and higher order effects~\cite{Berends:1973fd,Brown:1973ji,Cashmore:1985vp}. The forward-backward asymmetry is an odd function of the cosine of the polar angle in the center of mass system, $\cos\theta^*$. Since the acceptances of the Belle and Belle II detectors are not the same for $\cos\theta^*>0$ and $\cos\theta^*<0$, a charge asymmetry in the production of a given species of charmed hadrons remains. In our measurement, however, we effectively count pairs of charmed hadrons: the signal \DzToKSKS and the other (oppositely flavored) charmed hadron of the event, which provides the tagging information. As a result, we expect a negligible effect from the forward-backward asymmetry. To verify this, we weight the reconstructed candidates so that the $|\cos\theta^*|$ distribution of the signal is the same for candidates with $\cos\theta^*>0$ and $\cos\theta^*<0$ and redetermine the values of \Acp. As expected, we find variations in \Acp consistent with zero and do not assign any systematic uncertainty due to the forward-backward asymmetry.

Finally, as a cross-check we fit to subsamples of the data defined according to data-taking conditions and find no significant variation of the measured asymmetry.

The total systematic uncertainties, 0.43\% for Belle and 0.25\% for Belle II, are evaluated as the sums in quadrature of the components due to the fit modeling and the $\Dz\to\KS\pip\pim$ contamination.

\section{Final results and conclusions\label{sec:results}}
We measure the time-integrated \CP asymmetry in \DzToKSKS decays using a charm-flavor tagging algorithm that exploits the correlation between the flavor of the reconstructed neutral \D meson and the electric charges of particles reconstructed in the rest of the $\epem\to\ccbar$ event. Using 980\invfb of data collected by Belle and 428\invfb of data collected by Belle II, we obtain
\begin{align}
\Acp(\DzToKSKS) &= (\phantom{-}\resBelle\pm\statBelle\pm\systBelle)\%
\intertext{and}
\Acp(\DzToKSKS) &= (\resBelleII\pm\statBelleII\pm\systBelleII)\%\,,
\end{align}
respectively. The first uncertainties are statistical and the second systematic. The two results are in agreement and combined, using the best linear unbiased estimator~\cite{Valassi:2003mu}, into
\begin{equation}
\Acp(\Dz\to\KS\KS) = (\resValue\pm\resStat\pm\resSyst)\%\,.
\end{equation}
In the combination, the systematic uncertainties due to the fit modeling are considered uncorrelated, while those due to the $\Dz\to\KS\pip\pim$ contamination are considered fully correlated.

The results are also consistent with previous Belle and Belle II determinations based on the independent sample of \Dstarp-tagged \DzToKSKS decays~\cite{Belle:2024vho}. A combination of the result of this paper with that of Ref.~\cite{Belle:2024vho}, $(-1.4\pm1.3\pm0.1)\%$, yields
\begin{equation}
\Acp(\DzToKSKS) = (\resComb\pm\statComb\pm\systComb)\%\,.
\end{equation}
This is the most precise determination of $\Acp(\DzToKSKS)$ to date. It agrees with \CP symmetry and with results from other experiments~\cite{CLEO:2000opx,LHCb:2015ope,LHCb:2021rdn,CMS:2024hsv}.

 % end input ./body.tex
 
%
% start input ./acknowledgements.tex
%
This work, based on data collected using the Belle II detector, which was built and commissioned prior to March 2019,
and data collected using the Belle detector, which was operated until June 2010,
was supported by
Higher Education and Science Committee of the Republic of Armenia Grant No.~23LCG-1C011;
Australian Research Council and Research Grants
No.~DP200101792, %
No.~DP210101900, %
No.~DP210102831, %
No.~DE220100462, %
No.~LE210100098, %
and
No.~LE230100085; %
Austrian Federal Ministry of Education, Science and Research,
Austrian Science Fund (FWF) Grants
DOI:~10.55776/P34529,
DOI:~10.55776/J4731,
DOI:~10.55776/J4625,
DOI:~10.55776/M3153,
and
DOI:~10.55776/PAT1836324,
and
Horizon 2020 ERC Starting Grant No.~947006 ``InterLeptons'';
Natural Sciences and Engineering Research Council of Canada, Compute Canada and CANARIE;
National Key R\&D Program of China under Contract No.~2024YFA1610503,
and
No.~2024YFA1610504
National Natural Science Foundation of China and Research Grants
No.~11575017,
No.~11761141009,
No.~11705209,
No.~11975076,
No.~12135005,
No.~12150004,
No.~12161141008,
No.~12475093,
and
No.~12175041,
and Shandong Provincial Natural Science Foundation Project~ZR2022JQ02;
the Czech Science Foundation Grant No.~22-18469S 
and
Charles University Grant Agency project No.~246122;
European Research Council, Seventh Framework PIEF-GA-2013-622527,
Horizon 2020 ERC-Advanced Grants No.~267104 and No.~884719,
Horizon 2020 ERC-Consolidator Grant No.~819127,
Horizon 2020 Marie Sklodowska-Curie Grant Agreement No.~700525 ``NIOBE''
and
No.~101026516,
and
Horizon 2020 Marie Sklodowska-Curie RISE project JENNIFER2 Grant Agreement No.~822070 (European grants);
L'Institut National de Physique Nucl\'{e}aire et de Physique des Particules (IN2P3) du CNRS
and
L'Agence Nationale de la Recherche (ANR) under Grant No.~ANR-21-CE31-0009 (France);
BMBF, DFG, HGF, MPG, and AvH Foundation (Germany);
Department of Atomic Energy under Project Identification No.~RTI 4002,
Department of Science and Technology,
and
UPES SEED funding programs
No.~UPES/R\&D-SEED-INFRA/17052023/01 and
No.~UPES/R\&D-SOE/20062022/06 (India);
Israel Science Foundation Grant No.~2476/17,
U.S.-Israel Binational Science Foundation Grant No.~2016113, and
Israel Ministry of Science Grant No.~3-16543;
Istituto Nazionale di Fisica Nucleare and the Research Grants BELLE2,
and
the ICSC – Centro Nazionale di Ricerca in High Performance Computing, Big Data and Quantum Computing, funded by European Union – NextGenerationEU;
Japan Society for the Promotion of Science, Grant-in-Aid for Scientific Research Grants
No.~16H03968,
No.~16H03993,
No.~16H06492,
No.~16K05323,
No.~17H01133,
No.~17H05405,
No.~18K03621,
No.~18H03710,
No.~18H05226,
No.~19H00682, %
No.~20H05850,
No.~20H05858,
No.~22H00144,
No.~22K14056,
No.~22K21347,
No.~23H05433,
No.~26220706,
and
No.~26400255,
and
the Ministry of Education, Culture, Sports, Science, and Technology (MEXT) of Japan;  
National Research Foundation (NRF) of Korea Grants
No.~2016R1-D1A1B-02012900,
No.~2018R1-A6A1A-06024970,
No.~2021R1-A6A1A-03043957,
No.~2021R1-F1A-1060423,
No.~2021R1-F1A-1064008,
No.~2022R1-A2C-1003993,
No.~2022R1-A2C-1092335,
No.~RS-2023-00208693,
No.~RS-2024-00354342
and
No.~RS-2022-00197659,
Radiation Science Research Institute,
Foreign Large-Size Research Facility Application Supporting project,
the Global Science Experimental Data Hub Center, the Korea Institute of
Science and Technology Information (K24L2M1C4)
and
KREONET/GLORIAD;
Universiti Malaya RU grant, Akademi Sains Malaysia, and Ministry of Education Malaysia;
Frontiers of Science Program Contracts
No.~FOINS-296,
No.~CB-221329,
No.~CB-236394,
No.~CB-254409,
and
No.~CB-180023, and SEP-CINVESTAV Research Grant No.~237 (Mexico);
the Polish Ministry of Science and Higher Education and the National Science Center;
the Ministry of Science and Higher Education of the Russian Federation
and
the HSE University Basic Research Program, Moscow;
University of Tabuk Research Grants
No.~S-0256-1438 and No.~S-0280-1439 (Saudi Arabia), and
Researchers Supporting Project number (RSPD2025R873), King Saud University, Riyadh,
Saudi Arabia;
Slovenian Research Agency and Research Grants
No.~J1-50010
and
No.~P1-0135;
Ikerbasque, Basque Foundation for Science,
State Agency for Research of the Spanish Ministry of Science and Innovation through Grant No. PID2022-136510NB-C33, Spain,
Agencia Estatal de Investigacion, Spain
Grant No.~RYC2020-029875-I
and
Generalitat Valenciana, Spain
Grant No.~CIDEGENT/2018/020;
the Swiss National Science Foundation;
The Knut and Alice Wallenberg Foundation (Sweden), Contracts No.~2021.0174 and No.~2021.0299;
National Science and Technology Council,
and
Ministry of Education (Taiwan);
Thailand Center of Excellence in Physics;
TUBITAK ULAKBIM (Turkey);
National Research Foundation of Ukraine, Project No.~2020.02/0257,
and
Ministry of Education and Science of Ukraine;
the U.S. National Science Foundation and Research Grants
No.~PHY-1913789 %
and
No.~PHY-2111604, %
and the U.S. Department of Energy and Research Awards
No.~DE-AC06-76RLO1830, %
No.~DE-SC0007983, %
No.~DE-SC0009824, %
No.~DE-SC0009973, %
No.~DE-SC0010007, %
No.~DE-SC0010073, %
No.~DE-SC0010118, %
No.~DE-SC0010504, %
No.~DE-SC0011784, %
No.~DE-SC0012704, %
No.~DE-SC0019230, %
No.~DE-SC0021274, %
No.~DE-SC0021616, %
No.~DE-SC0022350, %
No.~DE-SC0023470; %
and
the Vietnam Academy of Science and Technology (VAST) under Grants
No.~NVCC.05.12/22-23
and
No.~DL0000.02/24-25.

These acknowledgements are not to be interpreted as an endorsement of any statement made
by any of our institutes, funding agencies, governments, or their representatives.

We thank the SuperKEKB team for delivering high-luminosity collisions;
the KEK cryogenics group for the efficient operation of the detector solenoid magnet and IBBelle on site;
the KEK Computer Research Center for on-site computing support; the NII for SINET6 network support;
and the raw-data centers hosted by BNL, DESY, GridKa, IN2P3, INFN, 
PNNL/EMSL, 
and the University of Victoria.
 % end input ./acknowledgements.tex
 
\bibliographystyle{belle2}
\providecommand{\href}[2]{#2}\begingroup\raggedright\endgroup

\end{document}